\documentclass[trackchanges]{aastex61}
\usepackage{graphicx}
\usepackage{amsmath}
\usepackage{amssymb}
\usepackage{wasysym}
\usepackage{pifont}

\usepackage[normalem]{ulem}
\usepackage{cancel}
\usepackage{color}

\newcommand{\vv}[1]{\mathbf{#1}}
\newcommand{\ds}{\displaystyle}

\begin{document}

\title{Chaotic zones around rotating small bodies}
\shorttitle{Chaotic zones around rotating small bodies}
\shortauthors{J. Lages, D.L. Shepelyansky, I.I. Shevchenko}

\author[0000-0001-5965-8876]{Jos\'e Lages}
\correspondingauthor{Jos\'e Lages}
\email{jose.lages@utinam.cnrs.fr}
\affiliation{Institut UTINAM, Observatoire des Sciences de l'Univers THETA, CNRS, Universit\'e de Franche-Comt\'e, Besan\c{c}on 25030, France}
\author[0000-0002-2752-0765]{Dima L. Shepelyansky}
%\email{dima@irsamc.ups-tlse.fr}
\affiliation{Laboratoire de Physique Th\'eorique du CNRS, IRSAMC, Universit\'e de Toulouse, UPS, Toulouse 31062, France}
\author[0000-0002-9706-0557]{Ivan I. Shevchenko}
%\email{iis@gao.spb.ru}
\affiliation{Pulkovo Observatory, RAS, 196140 Saint Petersburg, Russia}
\affiliation{Lebedev Physical Institute, RAS, 119991 Moscow, Russia }
\affiliation{Institut UTINAM, Observatoire des Sciences de l'Univers THETA, CNRS, Universit\'e de Franche-Comt\'e, Besan\c{c}on 25030, France}

\date{\today}

\begin{abstract}
Small bodies of the Solar system, like asteroids, trans-Neptunian objects, cometary nuclei, planetary satellites, with diameters smaller than one thousand kilometers usually have irregular shapes, often resembling dumb-bells, or contact binaries. The spinning of such a gravitating dumb-bell creates around it a zone of chaotic orbits. We determine its extent analytically and numerically. We find that the chaotic zone swells significantly if the rotation rate is decreased; in particular, the zone swells more than twice if the rotation rate is decreased ten times with respect to the ``centrifugal breakup'' threshold. We illustrate the properties of the chaotic orbital zones in examples of the global orbital dynamics about asteroid 243~Ida (which has a moon, Dactyl, orbiting near the edge of the chaotic zone) and asteroid 25143~Itokawa.
\end{abstract}


\section{Introduction}

The orbital dynamics around irregularly-shaped bodies (having
complex gravity fields) was extensively studied in the last two
decades. The reason is twofold: first, satellites of small bodies
such as asteroids were discovered; second, space missions were
planned and accomplished to asteroids and cometary nuclei.
Therefore, many aspects of the orbital dynamics in rotating
complex gravity fields were studied, both theoretically and in
numerical simulations; see \cite{Sch94,Sch12} and references
therein. Small bodies of the Solar system (asteroids,
trans-Neptunian objects, cometary nuclei, planetary satellites)
with diameters less than one thousand kilometers usually have
strongly irregular shapes \citep[p.~270]{melnikov10,jorda16}, in
many cases resembling dumb-bells, or ``contact binaries''. Various
models for gravity fields of the ``central body'' were used: that
of a triaxial ellipsoid with uniform density
\citep{CFM93,MOA06,O06,MA07}, a rod \citep{BB03}, a dumb-bell or
``bilobed'' model \citep{MDC14,FNV16}, a collection (``molecule'')
of gravitating points \citep{petit97}, a polyhedral model
\citep{W94,WS96}, a truncated gravitational field derived from a shape model \citep{feng17}. Orbits around actual small bodies, such as
asteroids Castalia, Eros, and Hektor were extensively modeled
\citep{SOH96,SWM00,MDC14,YB12}.  Concerning the dumb-bell model,
it was also used in the problem on spin-spin resonances in a
system of two aspherical gravitating bodies \citep{H81,BM15}: the
quadrupole moment of the secondary was represented as a dumb-bell
of two equal masses. This model provides  a setting for a
qualitative description of the tidal evolution and the resulting
spin-spin coupling of tight binary systems of elongated bodies
\citep{BM15}.

Many studies were devoted to resonant phenomena and determination
of orbital stability regions; see, in particular,
\cite{Sch94,HS04,MOA06,O06,MA07,Sch12} and references therein. The
existence of ``chaotic gravitational zones'' around rotating
elongated bodies was outlined by \cite{MOA06}. A destabilizing
role of resonances between particle's orbital motion and the
rotational motion of the central elongated body was revealed by
\cite{MOA06,O06,MA07}; in particular, see figures 1--5 in
\cite{O06} and figures 2 and 9 in \cite{MA07}, where the integer
spin-orbit resonances form a characteristic ``saw of instability''
in the plane of initial values of the semimajor axis and
eccentricity of the orbiting particle. Quite recently, numerical
simulations of orbits around contact binaries were performed by
\cite{FNV16} emphasizing the stabilization effect of the fast
rotation of the contact binary: for faster binary's rotation,
periodic orbits start to loose stability closer to the barycenter;
this is explained by averaging of the perturbation.

The preceding theoretical studies were based on the analysis of
perturbation functions and their expansions, in particular,
resonant terms in the expansions. In our article, we present a
different technique, based on  analysis of symplectic maps
\citep[see eg][]{meiss92}, in particular a generalized Kepler
map. The Kepler map approach allows one to understand
straightforwardly the global orbital behavior. Let us recall that
the Kepler map is a two-dimensional area-preserving map,
describing the eccentric circumbinary motion of a massless
particle in the gravitational field of a primary and a perturber
(the secondary moving around the primary in a circular orbit
deeply inside the particle's orbit). The motion is described in
terms of changes in particle's energy and conjugated orbital phase
measured at its apocenter and pericenter passages. In particular,
it was shown that the Kepler map describes the dynamics of
highly-eccentric comets \citep{petrosky86,malyshkin99}, Comet
Halley among them \citep{chirikov89}. In an appropriate physical
model, it explains the phenomenon of strong microwave ionization
of excited hydrogen atoms \citep{casati88} and autoionization of
molecular Rydberg states \citep{benvenuto94}. A review of the
Kepler map theory in a historical context is given in
\cite{shevchenko11}. Recent applications of the Kepler map theory
along with its corresponding advancements concern processes of
disintegration of three-body systems and Lev\'y flight statistics
in these processes \citep{shevchenko10}, capture of dark matter by
the Solar system and by binary stellar systems
\citep{lages13,rollin15a}, accurate symplectic map description of
the long-term dynamics of Comet Halley \citep{rollin15b}.
In this article,  the  Kepler map is used mostly for analytical purposes, so that to provide an analytical description of resonances and borders of dynamical chaos in the stability diagrams. However, it is also used as a numerical tool, whose advantage is in the enormously high speed of computation, which allows one to construct the stability diagrams with very high spatial resolution (see section \ref{stability}, Figs.~\ref{stabdiag},\ref{zone},\ref{idaito}). 

In our work, we consider a passively gravitating particle orbiting
a gravitating dumb-bell. If the dumb-bell is fixed in space, the
particle cannot gain or loose orbital energy or angular momentum
for its orbital motion, because their source is absent. But if the
dumb-bell rotates, the particle's energy or angular momentum may
vary strongly, so that the particle may even escape or fall on the
primary, depending on initial conditions. Obviously, one expects
that the particles close to the primary are more prone to such
disturbances than those away from it.

\begin{figure}[htb]
\centering
\includegraphics[width=0.5\columnwidth]{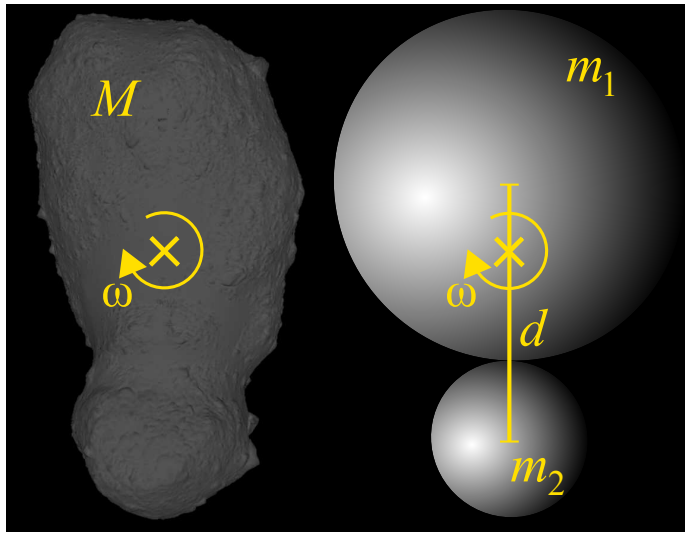}
\caption{\label{dumbbell}We model a non-axisymmetric small body
\citep[e.g. here 25143~Itokawa][]{gaskell08} by a contact binary
$m_1\geq m_2$ ($M=m_1+m_2$). The size of the dumb-bell is $d$, the
small body center of mass is marked by the cross. The axis of
rotation is perpendicular to the figure plane and passes through
the center of mass.}
\end{figure}

It is already known that a gravitating binary, such as a binary
star or a binary asteroid, has a circumbinary chaotic zone, where
all circumbinary orbits of the orbiting particles with any initial
eccentricity are chaotic \citep{shevchenko15}. But what would be
the case if one considers the motion around a rigid dumb-bell, for
which the spinning frequency $\omega$ can be smaller or larger
than the Keplerian frequency $\omega_0$ fixed by Kepler's third
law? Here we give an answer to this question generalizing the
Kepler map description \citep{chirikov89,petrosky86} to describe
the motion of a particle in the gravitational field of a spinning
body modeled by a dumb-bell with masses $m_1, m_2$ separated by
constant distance (dumb-bell size) $d$. In such a way, we model an
irregular body by two contact uniform-density spheres (equivalent
to two point masses) as it is shown in Fig.~\ref{dumbbell} for an
example of asteroid 25143 Itokawa \citep{gaskell08}. The dumb-bell
is spinning around its center of mass with an angular frequency
$\omega$, which can be different from the Keplerian frequency
$\omega_0$ of revolution of masses $m_1, m_2$. The dynamics of
particles orbiting the dumb-bell is considered in the plane
orthogonal to the spin axis.

The Kepler map description of orbits about a spinning
non-axisymmetric body is achieved by introducing a parameter,
$\omega$, which is the rate of rotation of the model contact
binary (see Fig.~\ref{dumbbell}). The value of $\omega$ is
arbitrary. We derive analytical expressions for the kick function,
representing the energy increment for the test particle when it
passes the apocenter of its orbit. We consider the planar case
{i.e.} the case of the orbits lying in the plane orthogonal to the
small body spin axis. We note that the Kepler map appears also for
molecular Rydberg states with a rotating dipole core
\citep{benvenuto94}. In the gravitational potential, the dipole
term cancels, and in the dumb-bell case the quadrupole and
octupole contributions of the central body's gravitational field
provide leading terms in the kick function. However we show that,
in a wide range of spinning frequencies $\omega<\omega_0$,
retaining the quadrupole term is enough to qualitatively describe
the chaotic zone around the spinning body. Strikingly, such a zone
swells significantly for $\omega < \omega_0$ down to a certain
threshold. In our approach we derive the kick function in a closed
form, valid in the whole range of parameters' values. To connect
our theoretical findings with observational data, we illustrate
the properties of the chaotic orbital zones in examples of the
global orbital dynamics about asteroid 243~Ida (which has a moon,
Dactyl, orbiting near the edge of the chaotic zone) and asteroid
25143~Itokawa.

\section{The Kepler map description}

We consider the motion of a passively gravitating particle
in the planar circular restricted three-body problem
$m_1$--$m_2$--particle, where the two masses $m_1$ and
$m_2$ are connected by a massless rigid rod, thus forming a
dumb-bell (see Fig.~\ref{dumbbell}).
The Keplerian rate of rotation of a contact binary, {i.e.} two tangent spheres, is
\begin{equation}
\omega_0 = \sqrt{\pi \mathcal{G} \rho / 3},
\label{om0_calc}
\end{equation}
where $\rho$ is the density of the irregular body \citep{scheeres07}.
For a typical density $\rho = 1$g$/$cm$^3$
we have $\omega_0 = 2.5 \times 10^{-4}$\,s$^{-1}$ corresponding to a period of about $7$ hours.
There are many observed asteroids with significantly larger rotation periods
\citep[see eg][]{pravec08}.
From now on we express the physical quantities in the following units:
$\mathcal{G}M=1$ (where $M=m_1+m_2$ is the total mass of the irregular gravitating body, we choose $m_2\leq m_1$ and we define $\mu=m_2/M\leq 0.5$),
$d=1$ is the size of the effective dumb-bell (Fig.~\ref{dumbbell}), and the Keplerian frequency
$\omega_0= \sqrt{\mathcal{G}(m_1+m_2)/d^3}=1$;
particle's energy per unit of mass, $E$, is then expressed in units
of $d^2 {\omega_0}^2$.
We consider solely the
case of prograde (with respect to the dumb-bell rotation) orbits
of the particle; analysis of the retrograde case is analogous.
The Kepler map for the motion around a gravitating dumb-bell, if one allows for the arbitrary rotation rate $\omega$ of the dumb-bell, takes the form \citep{casati88,benvenuto94}
\begin{equation}
\begin{array}{lll}
E_{i+1} = E_{i} + \Delta E\left(\phi_i\right),\quad
\phi_{i+1} &=& \phi_{i} + 2 \pi \omega/ \vert 2E_{i+1} \vert^{3/2}
\label{kmp}
\end{array}
\end{equation}
where the subscript $i$ enumerates the pericenter
passages with the rotation  phase $\phi_i = \omega t_i$ and the corresponding
particle energy $E_i$ taken at apocenter.
We retrieve the original Kepler map derived in \cite{chirikov89} and \cite{petrosky86} by setting $\omega=\omega_0=1$.
The equation for the rotation phase $\phi_i$ variation
is given by the third Kepler law. Originally the map has been derived for
particle's orbit with a period larger than the period
of the binary, ie $\omega/|2E|^{3/2} \leq 1$
\citep{casati88,benvenuto94}.
We obtain the kick function $\Delta E$ generalizing to the case of a binary with an arbitrary non-Keplerian rotation velocity $(\omega\neq\omega_0)$ the work of \citet{roy03} and \citet{heggie75} devoted to energy change in hard binary due to distant encounters. Defining $\vv{r}=\cos\left(\omega t\right)\vv{\hat a}+\sin\left(\omega t\right)\vv{\hat b}$ the position of the dumb-bell lobe of mass $m_2$ relative to the dumb-bell lobe of mass $m_1$ ($\vv{\hat a}$ and $\vv{\hat b}$ are any two orthogonal fixed directions of the plane), and $\vv R$ the position of the test particle relative to the barycenter of the two lobes, the equation of motion for the test particle around the dumb-bell is given by
\begin{equation}
\vv{\ddot{R}}=-\vv{\nabla_{\vv{R}}}\Phi\left(\vv R,\vv r,\mu,\omega\right)
\end{equation}
where the gravitational potential reads
\begin{equation}\label{phi}
 \Phi\left(\vv R,\vv r,\mu,\omega\right)=-\ds\frac{1-\mu}{\|\vv R+\mu\vv r\|}-\ds\frac{\mu}{\|\vv R-(1-\mu)\vv r\|}.
\end{equation}
Defining $r=\|\vv r\|$ and $R=\|\vv R\|$, the multipole expansion of the gravitational potential gives
\begin{equation}\label{gravpot}
 \Phi\left(\vv R,\vv r,\mu,\omega\right)
 =-\ds\frac{1}{R}-\mu\left(1-\mu\right)\ds\frac{r^2}{2R^3}\left(3\left(\ds\frac{\vv r\cdot\vv R}{rR}\right)^2-1\right)
  -\mu\left(1-\mu\right)\left(2\mu-1\right)\ds\frac{r^3}{2R^4}\left(5\left(\ds\frac{\vv r\cdot\vv R}{rR}\right)^3-3\ds\frac{\vv r\cdot\vv R}{rR}\right)
 +O\left(\ds\frac{r^4}{R^5}\right)
\end{equation}
Here, besides the $1/R$ term, the first two leading terms of the series are retained. This
turns out to be well enough for the purposes of the present
analysis, as comparisons of our results with previous simulations
show (see sections~\ref{ID} and \ref{IH}).
The energy increment
\begin{equation}\label{DEE}
 \Delta E(\mu,q,\omega,\phi)=-\int_{-\infty}^{+\infty} \vv{\dot R}\cdot\vv{\nabla}\left(\Phi+\ds\frac{1}{R}\right) dt
\end{equation}
of a test particle forced to follow a parabola the focus of which
is the dumb-bell barycenter is a function of the pericenter
distance $q$, and of the phase of the dumb-bell $\phi$ when the
test particle passes at pericenter. Here, the two lobes of the
rotating dumb-bell form a circular binary. Following \citet{roy03}
in the case of a circular binary but rotating at arbitrary
frequency rate $\omega$, keeping the two first leading terms for
the kick function (\ref{DEE}) we obtain
\begin{equation}\label{DE}
 \Delta E\left(\mu,q,\omega,\phi\right)
\simeq W_1\sin\left(\phi\right)+W_2\sin\left(2\phi\right).
\end{equation}
In equation (\ref{DE}), the exchange of energy between the small spinning body and the test particle is splat in two terms: the first harmonic comes from the octupole term ($\propto r^3/R^4$) of the gravitational potential multipole expansion (\ref{gravpot}) with amplitude
\begin{equation}\label{W1}
 W_1\simeq
\mu(1-\mu)(1-2\mu)2^{1/4} \pi^{1/2}
\omega^{5/2}
q^{-1/4} \exp \left( - \frac{2^{3/2}}{3} \omega q^{3/2} \right)
\end{equation}
and the second harmonic comes from the quadrupole term ($\propto r^2/R^3$) with amplitude
\begin{equation}\label{W2}
W_2\simeq
-\mu(1-\mu)2^{15/4} \pi^{1/2}
\omega^{5/2}
q^{3/4}\exp
\left( - \frac{2^{5/2}}{3}
\omega q^{3/2}
\right).
\end{equation}
We note that expression (\ref{W1}) restricted to the case $\omega=\omega_0=1$ and $\mu\ll1$ has been obtained using different method in \citet{shevchenko11}.
Usually, in the Kepler map ($\omega=\omega_0$) the kick function $\Delta E$
is proportional to $\sin \phi$  which is
just the first most prominent term in the Fourier expansion of the
energy increment, especially if $\mu \ll 1$ \citep{petrosky86,shevchenko11}.
This is for example the case when one consider the Kepler map description of cometary dynamics around the Solar System modelized by the Sun and Jupiter as perturber \citep{chirikov89,petrosky86}.
But with an increase of $\mu$ the second harmonic ($\propto \sin 2 \phi$)
becomes more and more important, and even remains the sole term for the case $\mu= 1/2$ since the first harmonic ($\propto \sin \phi$)
disappears ($W_1=0$). Indeed, for $\mu=1/2$, due to the
equality of the mass of primaries, by symmetry, the perturbation frequency is
effectively doubled.

\begin{figure}[t]
\centering
\includegraphics[width=0.32\columnwidth]{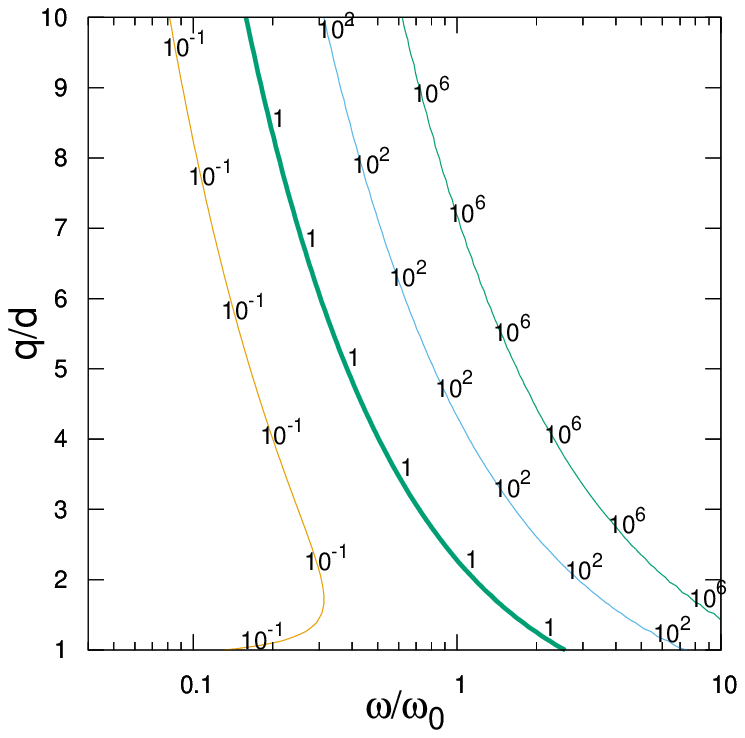}
\includegraphics[width=0.32\columnwidth]{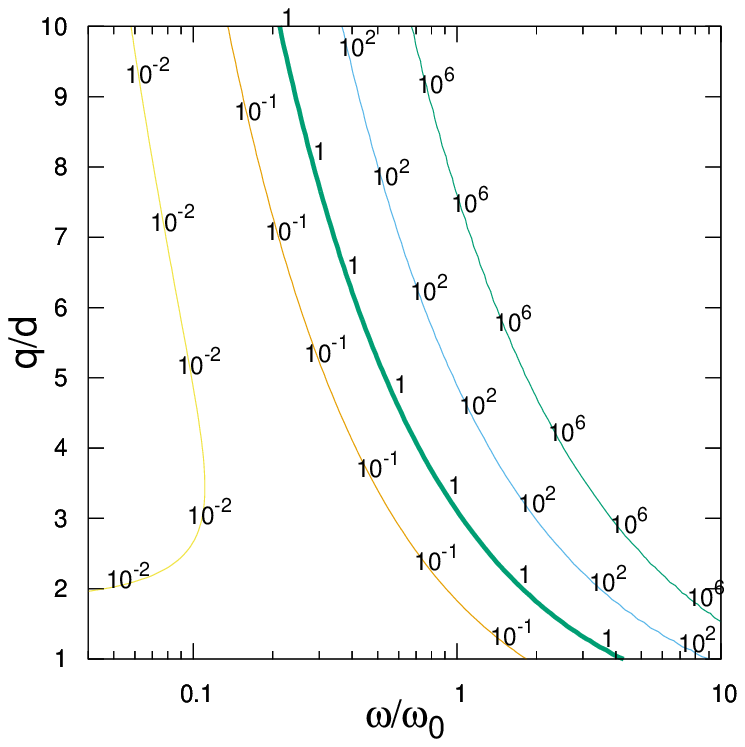}
\includegraphics[width=0.32\columnwidth]{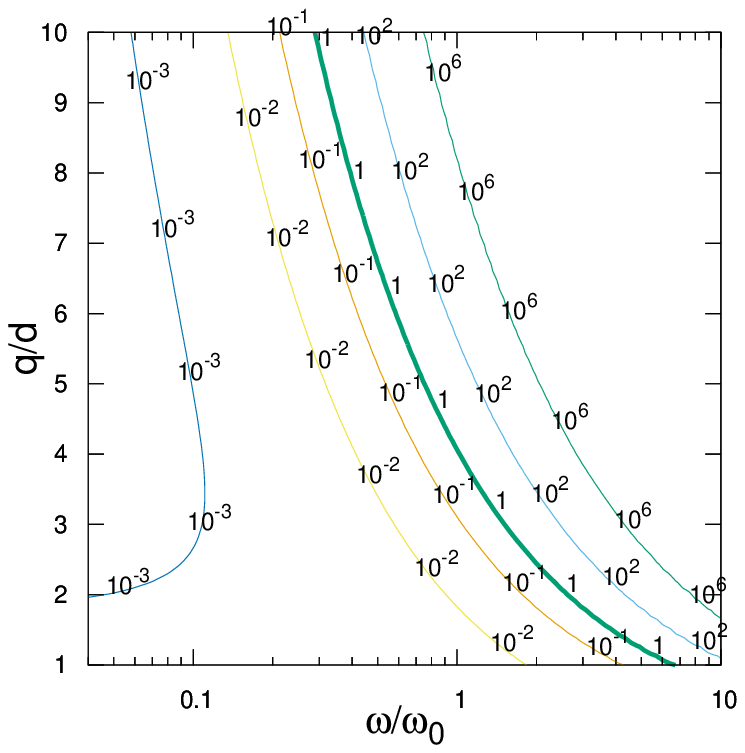}
\caption{\label{W1sW2} Contours of the function $W_1(\omega,q)/W_2(\omega,q)$
for
$\mu\rightarrow0$ (left panel),
$\mu=0.4$ (middle panel), and
$\mu=0.49$ (right panel).
}
\end{figure}

Here for the case of spinning small bodies a wide range of rotation frequencies can be considered; in particular spinning frequencies for asteroids range from $\omega/2\pi\sim10^{-3}$h$^{-1}$ to $\omega/2\pi\sim10^{2}$h$^{-1}$ \citep{whiteley02,warner09,hergenrother11}.
For $q\gg(\omega_0/\omega)^{2/3}d$, the contribution $W_1$ is obviously dominant since a factor $2$ exists between the arguments of the exponentials entering equations (\ref{W1}) and (\ref{W2}). This absolute prominence of $W_1$ over $W_2$ is even quadratically shifted farther from the small body for $\omega<\omega_0$.
Conversely, which contribution, either $W_1$ or $W_2$, dominates is not so obvious for the region $q\lesssim(\omega_0/\omega)^{2/3}d$ which for $\omega<\omega_0$ encompasses the immediate vicinity of the spinning small body.
The two contributions $W_1$ (\ref{W1}) and $W_2$ (\ref{W2}) depend on the parameters $\mu$, $\omega$ and $q$; their relative importance is summarized in the $(\omega,q)$ plot for different values of $\mu$ (Fig.~\ref{W1sW2}). We clearly see that below the frequency of disruption of a rubble-pile object ($\omega<\omega_0$), for any mass parameter $\mu$, the quadrupole coefficient $W_2$ generally dominates the octopole coefficient $W_1$ in the vicinity of the spinning small body. For example, $W_2\gg W_1$ for $q\lesssim3d$, $\omega\lesssim\omega_0$ , and for any $\mu$ parameter.

\begin{figure}[h]
\centering
\includegraphics[width=0.99\columnwidth]{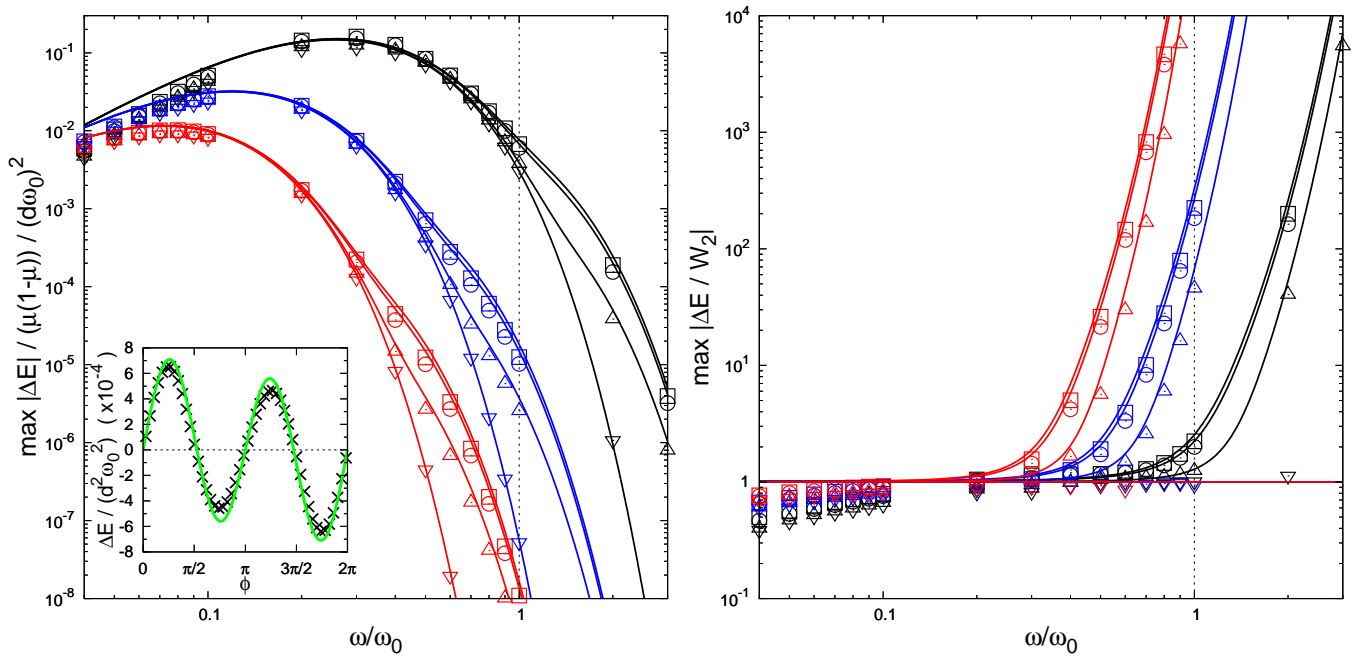}
\caption{\label{Famp}Amplitude of the energy kick $\Delta E$ as a function of the small body rotation frequency $\omega$,
computed by direct integration of the dynamics of a massless particle around a rotating dumb-bell,
for
\textcolor{black}{$q=3d$},
\textcolor{blue}{$q=5d$},
and
\textcolor{red}{$q=7d$},
and for different reduced masses
$\mu=0.01$ ($\square$),
$\mu=0.1$ ($\ocircle$),
$\mu=0.4$ ($\vartriangle$),
and
$\mu=0.5$ ($\triangledown$).
Plain lines give amplitudes of the analytically determined kick function $\Delta E$ (\ref{DE}) using (\ref{W1}) and (\ref{W2}).
Left panel: for the sake of clarity of the figure, amplitudes of the energy kick, $\underset{\phi}{\max}\arrowvert\Delta E\arrowvert$, are presented divided by the parameter $\mu(1-\mu)$.
Inset: example of kick function $\Delta E(\phi)$ for $q=5d$, $\mu=0.1$, and $\omega=0.3\omega_0$ computed from direct integration of the dynamics of a massless particle around a rotating dumb-bell (\ding{53}). The green solid line gives the kick function $\Delta E(\phi)$ (\ref{DE}).
Right panel: ratio $\underset{\phi}{\max}\arrowvert\Delta E/W_2\arrowvert$ with the same data as in the left panel.
}
\end{figure}

Typical amplitudes of energy kick functions $\Delta E$ are shown
in Fig.~\ref{Famp}. Analytical curves (\ref{DE}) constructed using
the first (\ref{W1}) and the second (\ref{W2}) harmonic terms of
the multipole expansion of the dumb-bell gravitational potential
are in good agreement with kick energy $\Delta E$ obtained by
direct integration of Newton's equations (Fig.~\ref{Famp}).
Globally the decrease of the small body spinning frequency induces
an increase of the energy kick. As expected for $q=7d$
insignificant kick ($\Delta E\sim10^{-8}d^2\omega_0^2$) is
expected in the case of an ordinary binary rotating with
$\omega=\omega_0$. However in the case of a spinning small body at
e.g. $\omega=0.1\omega_0$, the energy kick is strongly enhanced
($\Delta E\sim10^{-2}d^2\omega_0^2$). In comparison with ordinary
binary, such an energy kick increase induced by a slow spinning
frequency allows zone of chaos to extend quite far from the
central body. In Fig.~\ref{Famp} (left panel), amplitudes of kick
functions $\Delta E$ are presented divided by the mass factor
$\mu(1-\mu)$ entering the expression of $W_2$ (\ref{W2}). For
$q/d=3,5,7$ we clearly see that below
$\omega/\omega_0\simeq1,0.5,0.3$, curves for any reduced mass
$\mu$ are superimposed stressing again the fact that the second
harmonic term is dominant for small spinning frequencies (see also
Fig.~\ref{Famp}, right panel).

It should be noted that upon a minor modification this study
can be applied to a more generalized body, namely to a planar
molecule representing a set of coplanar asymmetric dumb-bells of
various size and $\mu$ with a common center of mass. In this way,
the Kepler map is straightforwardly generalized by means of adding
separate terms corresponding to each elementary dumb-bell's
contribution in the equation for the energy increment; each added
term has its own amplitude and constant phase shift in the body's
orientation.

In the frame of 3D atoms in a monochromatic field in 3D a
symplectic map was shown to give a correct description of real
dynamics (Casati et al. 1988). However, for a rotating gravitating
body, the generalization of our dumb-bell Kepler map to the 3D
case is an analytically complicated task, as a 3D generalization
of the classical Kepler map by \cite{emelyanenko90} shows. We
reserve this for a future study.

\section{Stability diagrams and central chaotic zone}
\label{stability}

\begin{figure}[t]
\centering
\includegraphics[width=0.99\columnwidth]{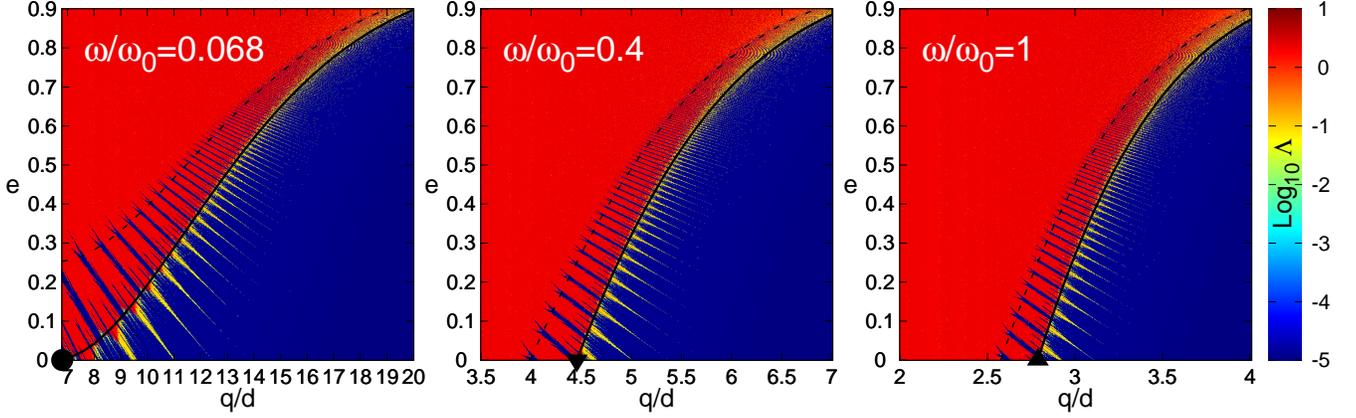}
\caption{\label{stabdiag}
Stability diagrams for $\mu=1/2$ and for $\omega/\omega_0=0.068$ (left panel), $0.4$ (middle panel), and $1$ (right panel). Chaotic (regular) domains are shown by reddish (blueish) areas. Chaos is
determined by computing the Lyapunov exponent $\Lambda$ for a
trajectory with initial orbital elements $(q,e)$. Here $10^6$ iterations of the Kepler map for dumb-bell (\ref{kmp}) have been computed for each initial orbital elements $(q,e)$.
The solid line gives the chaos border given by the analytical formula (\ref{Km_ecr}) with $K=K_G$. The dashed line gives the border of the bifurcation of half-integer resonances given by (\ref{Km_ecr}) with $K=2$.
Symbols \ding{108}, \ding{116}, and \ding{115} mark the limit of the central chaotic zone (see Fig.~\ref{zone}).
}
\end{figure}

Stability diagrams are constructed by computing Lyapunov exponents on a fine grid of initial data, $(e,q)$ or $(e,a)$. Lyapunov exponents are computed iterating concurrently the dumb-bell Kepler map (\ref{kmp}) and its tangent map (as, e.g., described by \cite{chirikov79} in application to the standard map).
The motion is regarded as
chaotic, if the maximum Lyapunov exponent is non-zero and
positive. Such diagrams are presented in the $(q,e)$ plane for
$\mu=1/2$ and for different values of $\omega/\omega_0=0.068$,
$0.4$, and $1$ (Fig.~\ref{stabdiag}). The border delimiting
chaotic domain (red) from regular domain (blue) is ragged; the
most prominent teeth being associated to the integer $p$:1 and
half-integer $p+\frac{1}{2}$:1 resonances between particle orbital
frequency and small body spinning frequency.
Here any neighboring
integer and half-integer resonances are equal-sized due to the
symmetry of the dumb-bell for $\mu=1/2$,
indeed
half-period and full-period rotations of the symmetric dumb-bell both result in configurations identical to the initial one.
The stability diagram
graphically demonstrates how the integer and half-integer
resonances overlap. Let us define the central chaotic zone as the
zone in $q$ such as at any initial eccentricity the particle's
dynamics is chaotic. Otherwise stated the chaotic zone is defined
as the region where even particles initially in circular orbits
become dynamically chaotic. From Fig.~\ref{stabdiag}, we clearly
see that the central chaotic zone swells significantly as the
small body spinning frequency decreases, since its farthest extent
varies from $q\simeq2.8d$ for $\omega=\omega_0$ to $q\simeq7d$ for
$\omega\simeq0.068\omega_0$.

Based on the concept of the chaotic layer around the separatrix and using analytical
expressions for the classical Kepler map parameter, a strictly
analytical expression for the size of the central chaotic zone
around a gravitating binary can be derived \citep{shevchenko15}. In a
similar way, the size of the central chaotic zone
around a rotating gravitating dumb-bell can be analytically
estimated.
Let us retain in (\ref{DE}) only the second harmonic contribution, since $W_2$ clearly dominates over $W_1$ for small spinning frequencies ($\omega<\omega_0$), indeed from Eqs. (\ref{W1}) and (\ref{W2}), for $\omega \ll (q/d)^{-3/2}\omega_0\ll\omega_0$, we obtain $W_2/W_1 \approx 2^{7/2} q /(1-2\mu)$ which is greater than $10$ for $\mu=0$ and diverges as $\mu$ approaches $1/2$.
By the substitution $E=W_2\,y$ and $\phi = x/2$ the map  (\ref{kmp})
is reduced to
\begin{equation}
y_{i+1} = y_i + \sin x_i,\qquad
x_{i+1} = x_i + \lambda/ \vert y_{i+1} \vert^{3/2}
\label{kmp_gm1}
\end{equation}
with $\lambda = 2^{1/2} \pi \omega /\vert W_{2} \vert^{3/2}$.
Following the standard procedure \citep{chirikov79,lichtenberg92,casati88}
the phase equation in (\ref{kmp_gm1}) can be linearized in $y$
in a vicinity of resonant phases $x=2\pi j$ with integer $j$
describing the local dynamics by the Chirikov standard map
with the chaos border
$y_\mathrm{cr}=(3\lambda/2K)^{2/5}$.
The
chaos parameter $K = K_\mathrm{G}=0.9716 \dots$ corresponds to the
critical golden curve \citep{lichtenberg92}. At $K > K_\mathrm{G}$, the
dynamical chaos is global, and the chaotic diffusion from
resonance to resonance becomes possible \citep{chirikov79,lichtenberg92}. However,
at $K$ exceeding $K_\mathrm{G}$ only slightly, relatively large
islands of stability exist inside the global domain of chaos.
At $K = 2$ bifurcation of half-integer resonances occur. At
this value the stability islands start to disappear.
The chaos border in energy is consequently
\begin{equation}
\Delta E_\mathrm{cr} = \left| W_{2} \,y_\mathrm{cr}  \right|
\approx A
\,\omega^{7/5}
q^{3/10} \exp \left(- B
\,\omega\,
q^{3/2}\right) , \label{DEcr}
\end{equation}
where $A = \mu^{2/5}\left(1-\mu\right)^{2/5}2^{13/10} 3^{2/5} \pi^{3/5} K^{-2/5}$ and $B= 2^{7/2} / 15$.
The half-width of the chaotic layer, $\Delta E_\mathrm{cr}$, and consequently the chaos border, is qualitatively well described by this Chirikov's criterion derived formula (see Fig.~\ref{ps} as an illustrative example).
The particle critical eccentricity $e_\mathrm{cr}$, following from
the relation $\Delta E_\mathrm{cr} = - E_\mathrm{cr} = 1/2 a_\mathrm{cr} =
(1 -
e_\mathrm{cr})/2 q$, is
\begin{equation}
e_\mathrm{cr} = 1 - 2 q \Delta E_\mathrm{cr} , \label{Km_ecr}
\end{equation}
where $\Delta E_\mathrm{cr}$ is given by
(\ref{DEcr}).

\begin{figure}[htb]
\centering
\includegraphics[width=0.49\columnwidth]{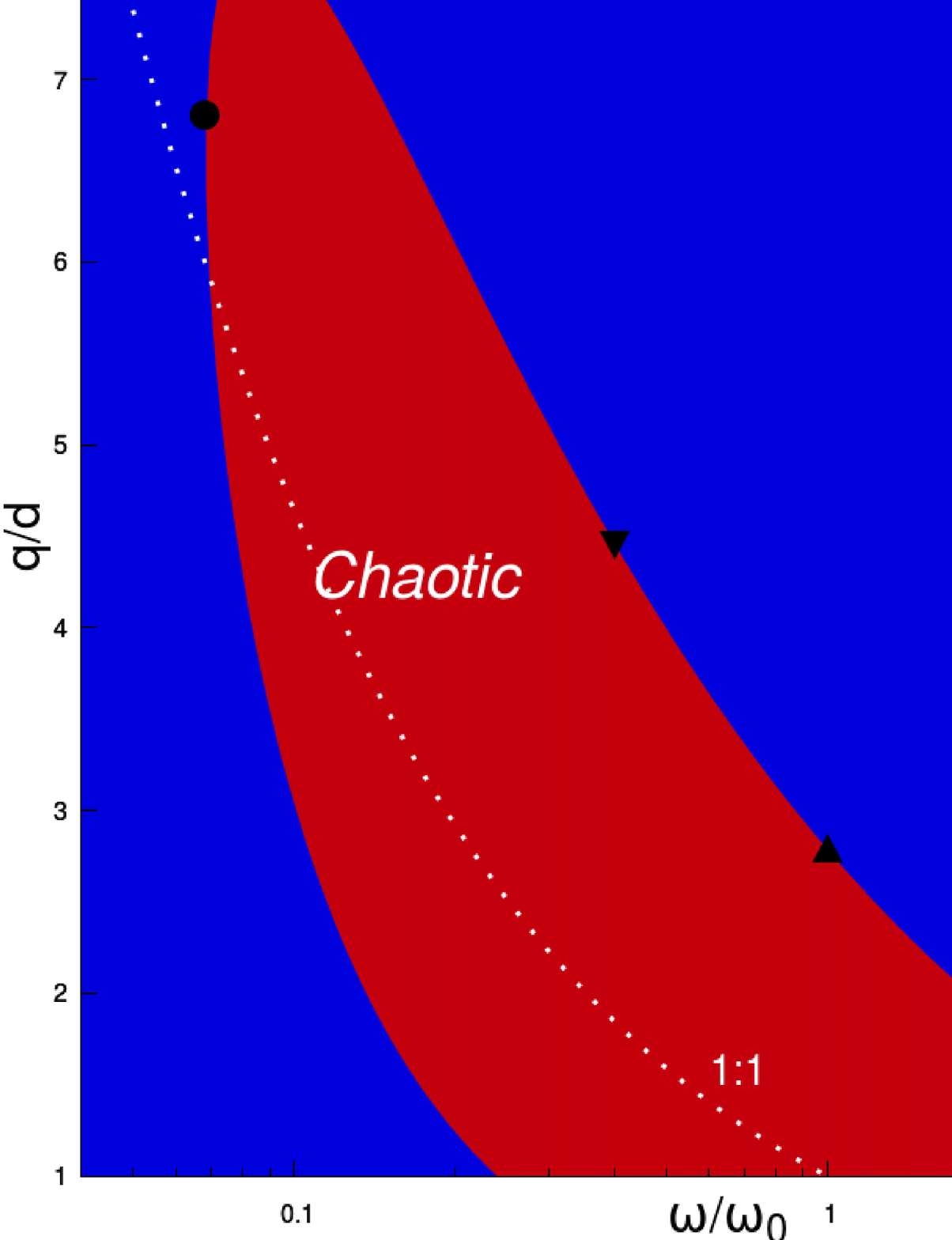}
\includegraphics[width=0.49\columnwidth]{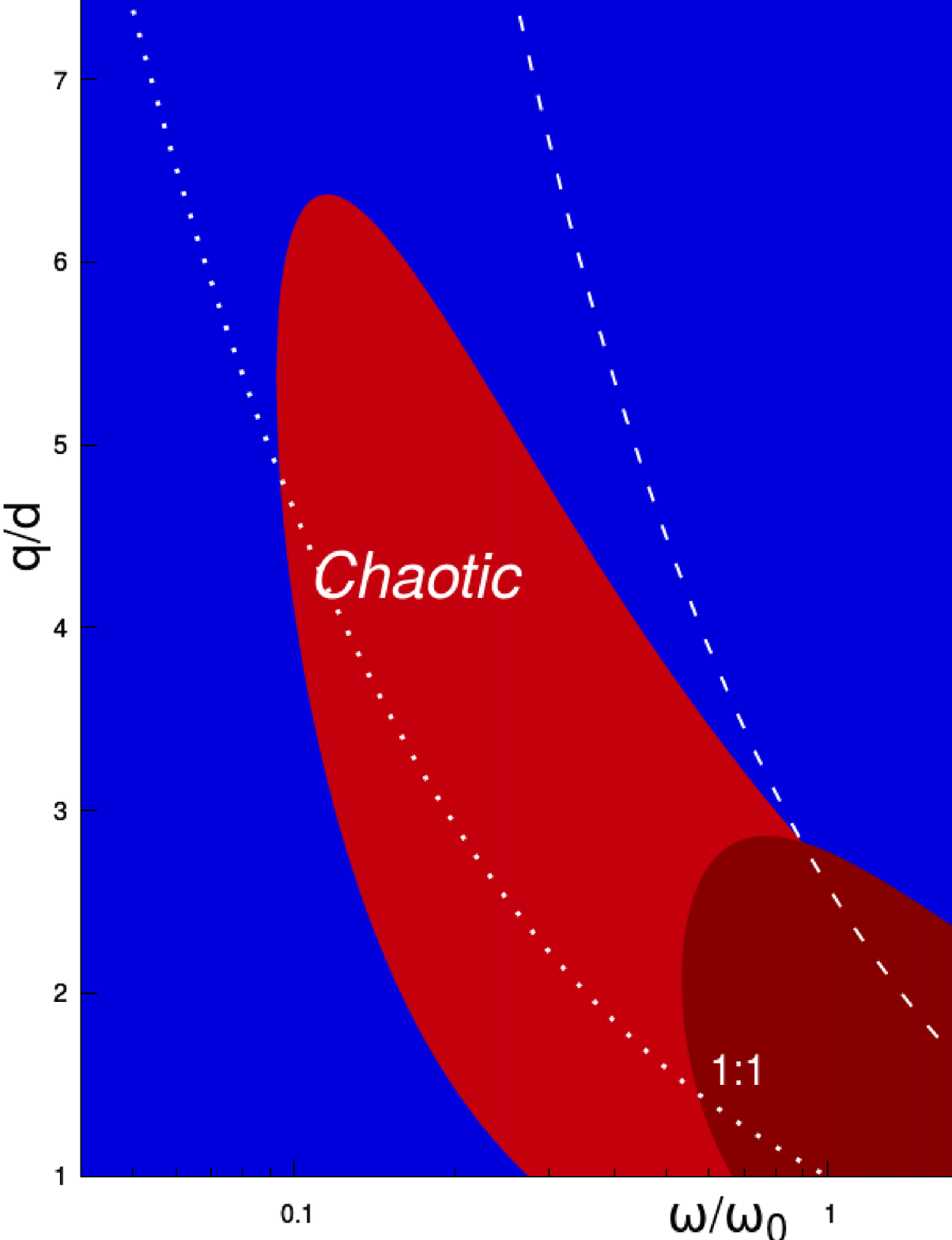}
\includegraphics[width=0.49\columnwidth]{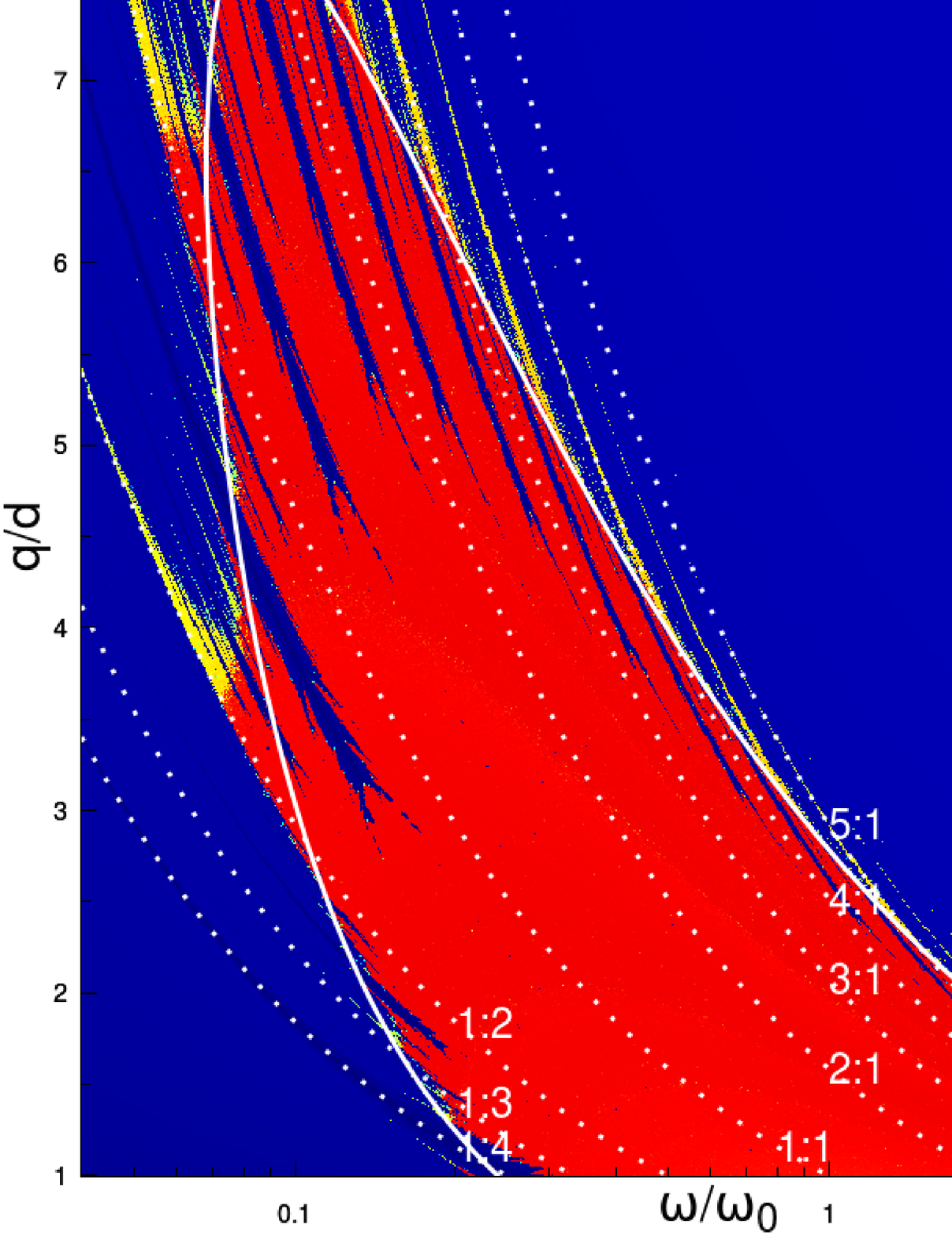}
\includegraphics[width=0.49\columnwidth]{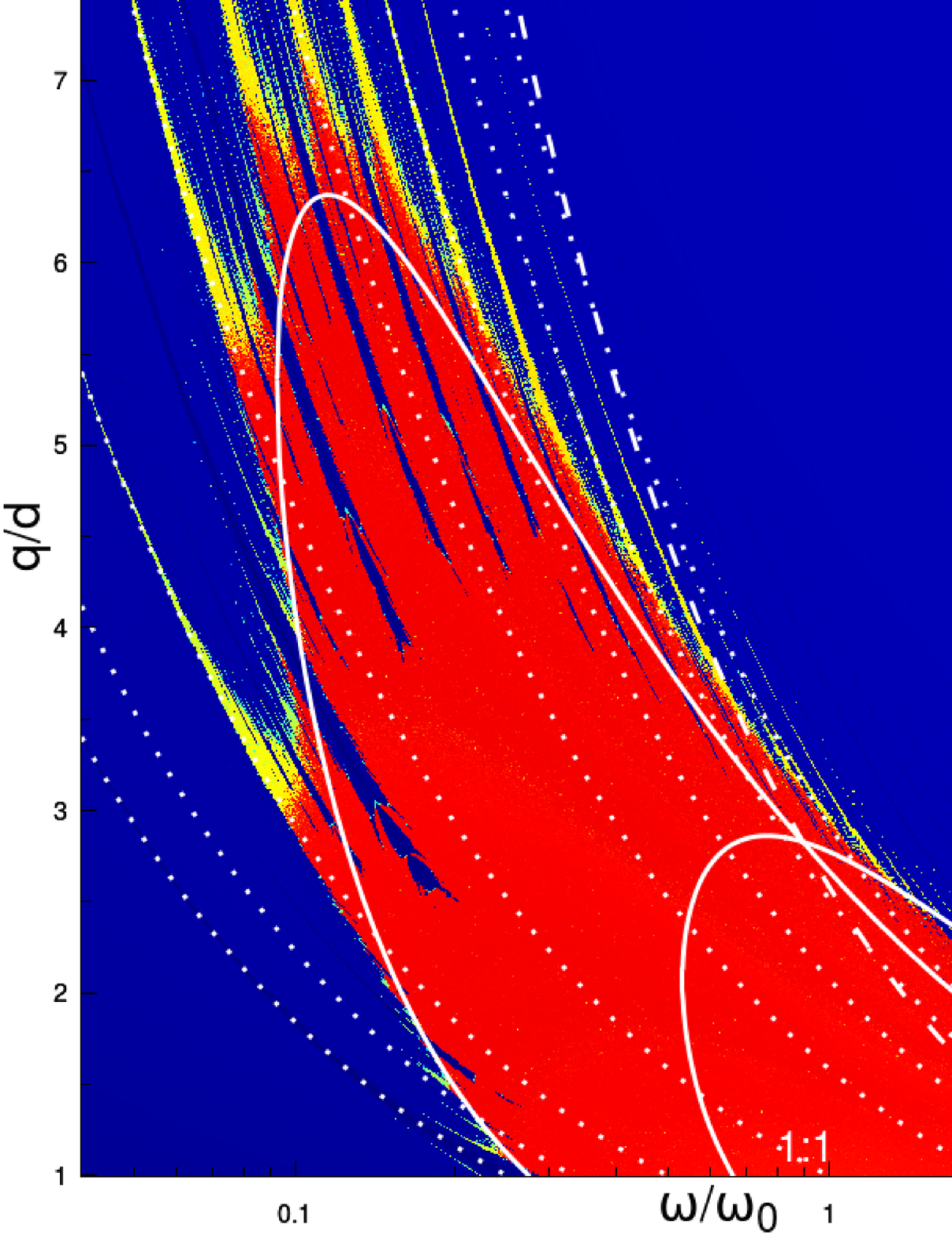}
\caption{\label{zone}Extent of the central chaotic zone around a small body as a function of the its spinning frequency $\omega$.
Upper left panel, case of a symmetric dumb-bell ($\mu=1/2$): the analytically obtained central chaotic zone is represented by the red domain. The blue area represents the complementary zone of stable orbits.
The symbols \ding{108}, \ding{116}, and \ding{115} mark the limit of the central chaotic zone for $\omega\simeq0.068, 0.4$, and $1$, respectively (see corresponding symbols in Fig.~\ref{stabdiag}).
Upper right panel, case of a non symmetric dumb-bell (here $\mu=1/2-\sqrt{1/12}\simeq0.211$): the central chaotic zone in red (dark red) is obtained analytically assuming that the second (first) harmonic term in (\ref{DE}) with amplitude $W_2$ ($W_1$) is dominant. The white dashed line represent the curve on which $W_1(q,\omega)=W_2(q,\omega)$.
Bottom left and right panels: stability diagrams in the $(q,\omega)$ plane for $e=0$. The reduced mass is $\mu=0.5$ (bottom left panel) and $\mu=1/2-\sqrt{1/12}\simeq0.211$ (bottom right panel). Chaos is determined by computing the Lyapunov exponent $\Lambda$. Here $10^6$ iterations of the Kepler map for dumb-bell (\ref{kmp}) have been computed for each couple of initial parameters $(q,\omega)$ with $e=0$. Solid white lines delimit central chaotic zones obtained analytically (see upper panels).
On each panels, white dotted lines represent $p$:1 and 1:$p$ resonances.
For the sake of clarity, all the resonances, marked by dotted lines, are labeled only in the bottom left panel. The location and distribution of resonances are determined by the ratio of orbital period to dumbbell spinning period. The resonance $p'\!\!:p$ is given by the curve $q/d=\left( \omega p/\omega_0p'\right)^{-2/3}$.
}
\end{figure}

Let us first consider $K=K_{\mathrm{G}}$, {i.e.} the value from which chaos is global: orbits with $e \gtrsim e_\mathrm{cr}(\omega,
q)$ are chaotic. In Fig.~\ref{stabdiag}, the analytical curve $e_\mathrm{cr}(q)$, given by
(\ref{DEcr}) and (\ref{Km_ecr}) at $K=K_{\mathrm{G}}$, is superimposed on stability diagrams for different values of $\omega$.
One can see that the $e_\mathrm{cr}(q)$ curve (black solid line) approximately
describes the ragged border of the chaotic zone.
At $K=2$, {i.e.} the value at which bifurcation of half-integer resonances of the standard map occurs, the $e_\mathrm{cr}(q)$ curve is shown by black dashed line in Fig.~\ref{stabdiag}. This curve gives the location where regular islands are no more distinguishable. The good performance of the analytical expression of $e_\mathrm{cr}(q)$ for $K=K_\mathrm{G}$ and $K=2$ testifies the adequacy of the map's theoretical model \citep{popova16}.

By calculating the $e_\mathrm{cr}(\omega, q)$ dependence, given by
(\ref{Km_ecr}) at $K=K_\mathrm{G}$, one can find the limits $q_1(\omega)$ and $q_2(\omega)$ of the central chaotic
zone around the spinning irregular body; these limits ($q_1<q_2$) are the roots of the equation
$e_\mathrm{cr}(q)=0$ at $\omega$ fixed.
Trajectories with $q_1<q<q_2$ and any initial eccentricity are
chaotic.
In Fig.~\ref{zone}, upper left panel, the central chaotic zone around a spinning symmetric dumb-bell ($\mu=1/2$) is represented by the red domain.
This global picture confirms that the central chaotic zone swells significantly as $\omega$ decreases. For $\mu=1/2$ the farthest limit of the central chaotic zone, $q\simeq7.8d$, occurs for $\omega\simeq0.08\omega_0$. This is $\sim2.8$ times the farthest limit for the Keplerian frequency $\omega=\omega_0$. Conversely, the increase of $\omega$ beyond $\omega_0$ leads to a shrinking of the central chaotic zone in agreement with the stabilization effect around fast rotating contact binary \citep{FNV16}.

The swelling of the central chaotic zone can be explained analyzing the $\omega$ dependence of the kick amplitude $W_2$ (\ref{W2}) and of the width $2\Delta E_\mathrm{cr}$ (\ref{DEcr}) of the chaotic layer around the separatrix ($E=0$). Taking the example of a symmetric dumb-bell ($\mu=1/2$), for $q=5d$ and a spinning rate $\omega=\omega_0$, the kick amplitude,  $W_{2}\approx10^{-8}\left(d\omega_0\right)^2$ (see Fig.~\ref{Famp}, left panel), is inefficient to produce chaotic orbits at any eccentricity since the lowest reachable semi-major axis is $a_\mathrm{cr}=1/(2\Delta E_\mathrm{cr})\approx500d$ and the lowest reachable eccentricity is $e_\mathrm{cr}\approx0.99$. For $q=5d$, but with a much slower dumb-bell spinning rate e.g. $\omega=0.3\omega_0$, the kick amplitude is switched on, $W_{2}\approx2\cdot10^{-3}\left(d\omega_0\right)^2$ (see Fig.~\ref{Famp}, left panel), in comparison to the $\omega=\omega_0$ case, giving $a_\mathrm{cr}\approx q$, and thus creating a chaotic layer with orbits of any eccentricity.
As a remark we note that the swelling of the chaotic zone at $\omega < 1$ has some price: the Lyapunov exponent decreases being proportional to $\omega$ at $\omega/\omega_0 \ll 1$.

For $\omega\lesssim0.24\omega_0$ a central regular zone appears in the immediate vicinity of the irregular small body. This central regular zone is surrounded by the chaotic zone and
increases as $\omega$ is decreased from $\omega\simeq0.24\omega_0$ down to $\omega\simeq0.068\omega_0$. This central regular zone appears in a region where test particles with circular orbits have a period smaller than the rotation period of the irregular small body (see white dotted line for 1:1 resonance in Fig.~\ref{zone} upper left panel).
Below  $\omega\simeq0.068\omega_0$, no zeros of (\ref{Km_ecr}) exists and consequently no central chaotic zone exists around the irregular spinning body (i.e. for $q>1$).

The most extended chaotic zone is provided by the symmetric case ($\mu=1/2$, Fig.~\ref{zone}, upper left panel). 
For the opposite case, at $\mu$ tending to zero the chaotic zone vanishes, because the perturbation from the second (smaller) lobe tends to zero.
Hence for intermediary cases with $\mu<1/2$, the chaotic zone is less extended, and the octopole contribution $W_1$, though weak for small $\omega$'s, is not negligible around and beyond $\omega=\omega_0$. Fig.~\ref{zone}, upper right panel, gives the example of the central chaotic zone for non symmetric dumb-bell with $\mu=1/2-\sqrt{1/12}\simeq0.211$. We have computed the analytical border of the central chaotic zone using (\ref{Km_ecr}) with either, as explained above, only the second harmonic contribution $W_2$ (red domain in Fig.~\ref{zone}, upper right panel), or with only the first harmonic contribution $W_1$ instead of $W_2$ (dark red domain in Fig.~\ref{zone}, upper right panel).
We observe quite a continuous overlap between the chaotic zones induced by the two contributions $W_1$ and $W_2$ which give together a qualitative global picture of the central chaotic zone around an irregular spinning small body. The white dashed line represent the contour where $W_1=W_2$.
The dependence of $W_2$ in $\mu$ tells us that the chaotic domain induced by $W_2$ is less and less wide as $\mu$ decreases from $\mu=1/2$ toward $\mu=0$. The chaotic domain induced by $W_1$ is the widest for $\mu=1/2-\sqrt{1/12}\simeq0.211$, and is less and less wide as $\mu$ increases (decreases) from $\mu\simeq0.211$ toward $\mu=1/2$ ($\mu=0$). Fig.~\ref{zone}, bottom panels, show stability diagrams of test particles initially in circular orbits ($e=0$) for the symmetric case $\mu=1/2$ (left panel) and for the $\mu=1/2-\sqrt{1/12}\simeq0.211$ non-symmetric case (right panel). The fractal contour (Fig.~\ref{zone} bottom panels) of the central chaotic zone around the small body is well approximate by analytically obtained contours (\ref{Km_ecr}).

\section{Ida and Dactyl}
\label{ID}

\begin{figure}[h]
\includegraphics[width=0.99\columnwidth]{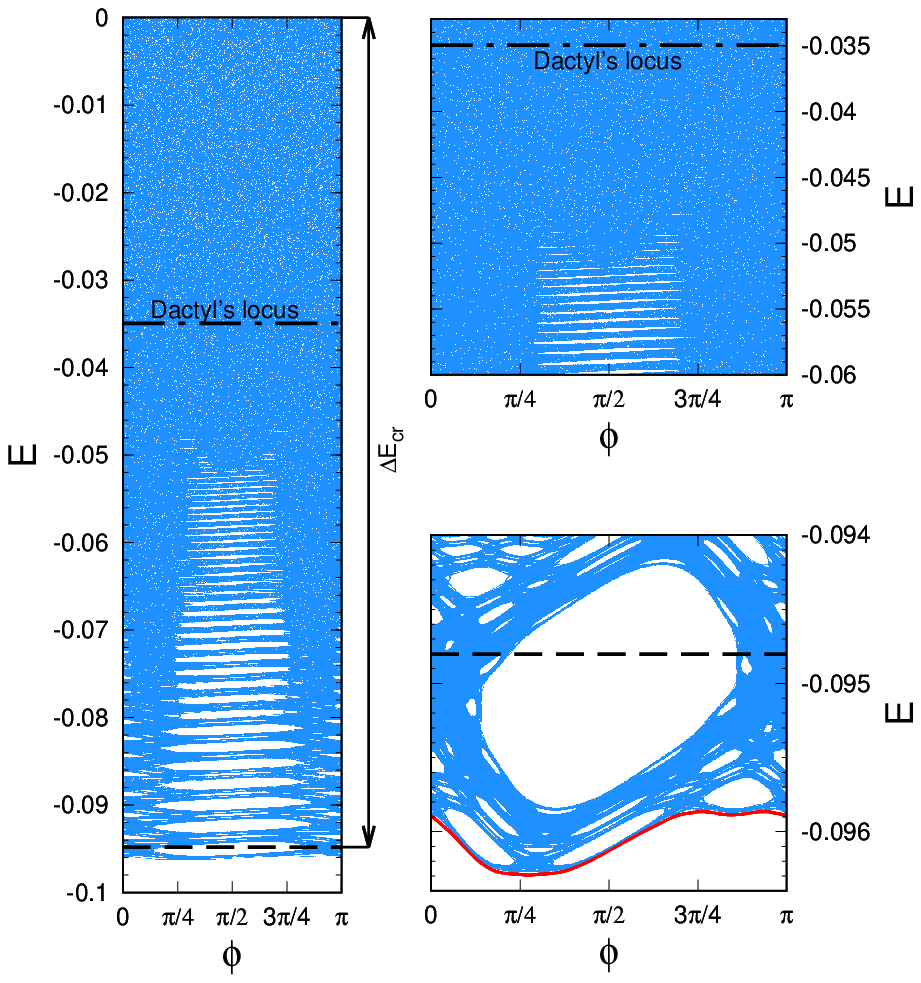}
\caption{\label{ps}  Poincar\'e section $(E,\phi)$ for Dactyl's dynamics around Ida computed from
(\ref{kmp}) with $\mu=1/2$. Left panel: the chaotic layer in the vicinity of the separatrix (chaotic sea) is shown by light
blue color. Ida's parameters $(d\simeq24.9\mathrm{km}$, $\omega \simeq0.953\omega_0)$ have been derived from physical parameters \citep{petit97,vokrouhlick03}.
A possible dynamical locus of Dactyl $(a\simeq14.3d$, $q\simeq3.20d)$, derived from \citep[][black point on Fig.~19]{petit97},
is shown by the dash-dotted line. The dashed line shows the analytical estimation of
the chaotic sea border according to Chirikov's criterion (\ref{DEcr}) at $K=K_\mathrm{G}$. Top right panel: close-up
around Dactyl's dynamical location. Bottom right panel: close-up of the chaotic sea border. The last invariant
 KAM curve separating the chaotic sea (above) from the regular domain (below) is shown in red.}
\end{figure}

\begin{figure}[h]
\includegraphics[width=0.99\columnwidth]{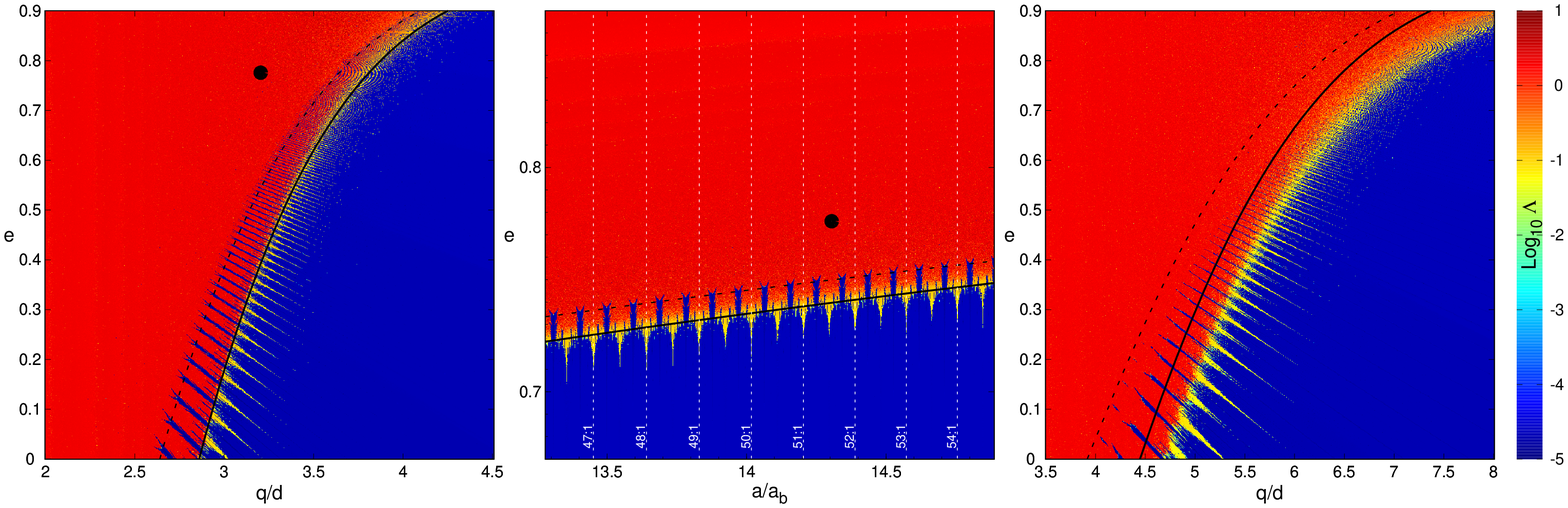}
\caption{\label{idaito} Stability diagram around Ida (left and
middle panels, $d\simeq24.9 \mathrm{km},\omega
\simeq0.953\omega_0,\mu\simeq1/2$) and around Itokawa (right
panel, $d\simeq280\mathrm{m},\omega
\simeq0.37\omega_0,\mu\simeq0.26$): the chaotic domain is shown by
the reddish area. Chaos is determined by computing the Lyapunov
exponent $\Lambda$ for a trajectory with initial orbital elements
$(q,e)$. Here $10^6$ iterations of the Kepler map for dumb-bell
(\ref{kmp}) have been computed for each initial orbital elements
$(q,e)$ (left and right panels) or $(a,e)$ (middle panel). The
black dot (left and middle panels) gives a possible current
dynamical position of Dactyl according to
\cite[][Fig.~19]{petit97}. The critical curves $e_\mathrm{cr}(q)$
(\ref{Km_ecr}) for overlap of integer resonances
$K=K_{\mathrm{G}}$ (solid line) and for bifurcation of
half-integer resonances $K=2$ (dashed line) are plotted.}
\end{figure}

We now apply the Kepler map approach to real celestial bodies.
Among the Solar system bodies, there exists quite a marked size
border line between the close-to-spherical large bodies and the
essentially ellipsoidal (potato-like) small bodies. This border
lies at $R=300$--500~km, where $R$ is the characteristic radius of
the body (see figures~1--2 in \cite{melnikov10}).

Moreover, usually asteroids and cometary nuclei resemble
dumb-bells, {i.e}., they are more like dumb-bells than ellipsoids.
A well-known example is the nucleus of comet
67P/Churyumov--Gerasimenko, the target of the Rosetta mission
\citep{jorda16}. Another example is asteroid 25143 Itokawa
(Fig.~\ref{dumbbell}), the target of the Hayabusa mission
\citep{fujiwara06}. In fact, several asteroids are observed to
have a bilobed shape; in particular, 243~Ida among them, is famous
to have a small natural satellite. The satellite, named Dactyl,
moves in an orbit prograde with the rotation of Ida, with a very
small inclination \citep[$i< 8^\circ$][]{petit97} with respect to
the equatorial plane of Ida.

Asteroid 243~Ida can be approximately described as a symmetric dumb-bell ($\mu=1/2$).
As follows from data presented in
\citet{belton95,belton96,petit97}, Ida resembles an aggregate of two
merged bodies with the ratio of masses $m_2 / m_1 \simeq 1$
\citep{petit97}.
We set the density $\rho$ and the rotation period $P_\mathrm{rot}
= 2 \pi / \omega$ of the asteroid, respectively, to be equal to
$2.24$~g.cm$^{-3}$ \citep{petit97} and $4.63$~h \citep{vokrouhlick03}.
% (Note that the mass of Ida in the same model is $M \simeq 3.6 \times 10^{19}$~g \citep{petit97}.)
Using formula~(\ref{om0_calc}), the corresponding spinning frequency for Ida is $\omega \simeq 0.953\omega_0$.
Besides, for the twin binary, consisting of two tangent spherical
masses $m$, one has $\rho \pi d^3 / 3 = 2m = M$, where $M$
and $d$ are, respectively, the total mass and size of the
dumb-bell. Therefore, for Ida one has $d \simeq 24.9$~km.

As an illustration (Fig.~\ref{ps}) we show the phase portrait
$(E,\phi)$ of Dactyl's dynamics around Ida obtained by iteration
of the Kepler map (\ref{kmp}). As discussed above, by calculating
the $e_\mathrm{cr}(\omega, q)$ dependence, given by (\ref{Km_ecr})
at $K=K_\mathrm{G}$, one can find the radius of the central
chaotic zone around the asteroid; it is given by the root of the
equation $e_\mathrm{cr}(q)=0$ at $\omega\simeq0.953\omega_0$. In
the case of Ida, the root is $q \simeq 2.85 d \simeq 71$~km. This
estimate for the chaotic zone extent is in good qualitative
agreement with the numerical-experimental findings on the
stability limit for Dactyl's orbit size found in \cite{petit97}.

Critical curves at $K=K_\mathrm{G}$ and at $K=2$ are superimposed
on stability diagrams for Ida in the $(q,e)$ plane
(Fig.~\ref{idaito}, left panel) and in the $(a,e)$ plane
(Fig.~\ref{idaito}, middle panel); the location of Dactyl is shown
by a black dot. The 51/1 and 52/1 resonant teeth engulf the cell
where Dactyl is located. The resonances densely accumulate higher
in the diagram, on approaching the parabolic separatrix. From
Figs.~\ref{idaito} it is clear that Dactyl is chaotic, in
agreement with the numerical-experimental findings in
\cite{petit97}.

Note that, in fact, short-time observations from the Galileo spacecraft gave no data on the stability of Dactyl's orbit. It can well be chaotic and thus short-lived. On the other hand, the determination of Dactyl's orbit may have also suffered inaccuracies (again due to the shortness of the observations), occasionally placing Dactyl in the chaotic region of the diagram. 

\section{Itokawa and Hayabusa}
\label{IH}

In the case of Ida, $\omega/\omega_0$ is not far from unity,
therefore, the found central chaotic zone is analogous to the one
existing usual Keplerian binary. In our second example,
25143~Itokawa, $\omega$ is much less than $\omega_0$ and chaotic
zone's swelling is expected to be large.

Itokawa was the target of the Hayabusa mission \citep{fujiwara06}.
Its shape is bilobed (Fig.~\ref{dumbbell}), and is described as a
contact binary of two ellipsoids with sizes
490$\times$310$\times$260~m (``body'') and
230$\times$200$\times$180~m (``head''), and densities
1750~kg/m$^3$ and 2850~kg/m$^3$, respectively; the centers of the
ellipsoids are separated by $d \simeq 280$~m~\citep{lowry14}. The
period of rotation of Itokawa is 12.132~h~\citep{alainen03}, and
its mass is estimated as $3.58 \times
10^{10}$~kg~\citep{fujiwara06}. Based on these observational data
one readily calculates: $P_0 = 2\pi / \omega_0 = 4.54$~h, $\omega
= 0.37$, $m_1/m_2 \simeq 2.9$, $\mu\simeq0.26$.

The stability diagram computed on the basis of these data using
the Kepler map for non-symmetric dumb-bell (\ref{kmp}) is shown in
Fig.~\ref{idaito}, right panel. The radius of the central chaotic
zone $q\simeq4.6$ is almost twice the chaotic zone we would have
obtained for Itokawa's parameters but $\omega=1$ (not shown). We
also clearly see that the central chaotic zone radius is well
estimated by the critical curves derived using only the second
harmonic contribution $W_2$.

Owing to the small mass, Itokawa's zone of gravitational influence
measured by its Hill radius $R_\mathrm{Hill}$ is also pretty
small: it can be as small as 25~km~\citep{fuse08}. What is more,
for a probe with large solar panels as Hayabusa, due to the effect
of the Solar radiation pressure the outer limits of the zone of
Itokawa's ability to sustain satellites diminish substantially to
about 3~km \citep{zimmer14}.

On the other hand, numerical modeling in \cite{zimmer14} showed
``that orbits below 1~km in semimajor axis were more susceptible
to the complex gravity of a rotating, non-uniform body with the
spacecraft either impacting or being ejected after only a few
orbits''. That is why, instead of trying to orbit Itokawa,
Hayabusa moved in a neighboring orbit around the Sun. From
Fig.~\ref{idaito}, right panel, it is clear that indeed no stable
circular orbit can be found below $q\simeq4.6d\simeq1.3$~km.

Itokawa has no satellites, as reported in \cite{fuse08}. The
formation of the extended central chaotic zone, in concert with
the smallness of the Hill sphere, explains the lack of moons. This
effect also explains why Hayabusa could not be put in orbit around
Itokawa.

\section{Capture cross-section}

Particles flying by a non spherical spinning body can be captured.
Following \cite{lages13}, the capture cross-section $\sigma$ characterizes the probability that a spinning body captures a scattering particle after a passage at the pericenter. The fact that chaotic zones increases significantly at $\omega/\omega_0 \ll 1$ leads
to an increase of the capture cross-section $\sigma$. Indeed, according to \cite{lages13} we have
$\sigma \sim \pi r_{st}^2 \sim qd(d \omega_0/v_{st})^2$ where
$r_{st}$ and $v_{st}$ are the impact distance and the mean velocity of a
scattering particle at infinity. Since from (\ref{W1})-(\ref{W2})
the exchange of energy (\ref{DE}) is non negligible for
pericenters up to $q \sim d (\omega_0/\omega)^{2/3}$ the above
estimate shows that the capture cross-section of slowly spinning
body $(\omega/\omega_0 \ll 1)$ can be significantly enhanced comparing to its geometric
cross-section $\sim\pi d^2$. Such an effect may play an important role for dust
capture by e.g. a spinning satellite.

\section{Conclusions}

We have generalized the Kepler map technique to describe the
motion of a particle in the gravitational field of a rotating
irregular body modeled by a dumb-bell. This has been achieved by
introduction of an additional parameter responsible for the
arbitrary rate of rotation of the ``central binary''. We have
found that the chaotic zone around the dumb-bell swells
significantly if its rotation rate is decreased; in particular,
the zone swells more than twice if the rotation rate is decreased
ten times with respect to the ``centrifugal breakup'' threshold.
We have determined the extent of the chaotic zone both
analytically and numerically.

To connect our theoretical findings with observational data, we
have illustrated the properties of the chaotic orbital zones in
examples of the global orbital dynamics about asteroid 243~Ida
(which has a moon, Dactyl, orbiting near the edge of the chaotic
zone) and asteroid 25143~Itokawa.

Possible orbital regimes of Ida's moon Dactyl have been described
by means of constructing stability diagrams of its orbital motion.
The qualitative dynamics of the  Ida--Dactyl asteroid--satellite
system has been shown to be described adequately within this
approach; in particular, an agreement has been found with previous
numerical-experimental data on the stability of orbits around Ida.
It has been explained why Dactyl is marginally chaotic, as its
orbit is situated at the fractal border of the analytically
expected central chaotic zone.

For Itokawa, it has been explained why space probe Hayabusa could
not be put in orbit around it, and also why Itokawa has no natural
satellites. All this is due to the swelling of the chaotic zone
around slowly rotating Itokawa, this enlargement being combined
with the smallness of its Hill sphere.

We highlight various important implications of emerged chaos
around rotating minor bodies. The dumb-bell map technique might be
perspectively applied to describe orbital motions and assess the
possibility and sizes of chaotic zones around elongated minor
planetary satellites, eg, minor moons in the Pluto--Charon
system. Indeed, as outlined in \cite{Q17}, in this system only
Hydra rotates rapidly (at the rate of $\sim$30\% of the
``centrifugal breakup'' threshold). Therefore, the chaotic zones
around the minor moons in the Pluto--Charon system may engulf
their Hill spheres substantially; this issue deserves further
study.

\acknowledgements

The authors are thankful to Jean-Marc Petit, Darin Ragozzine and anonymous referee for valuable remarks and comments.
I.I.S. benefited from a grant
of Bourgogne-Franche-Comt\'e region. I.I.S. was supported in
part by the Russian Foundation for Basic Research (project No.
17-02-00028).

\bibliography{CZAQRbib}{}

\begin{thebibliography}{}
\expandafter\ifx\csname natexlab\endcsname\relax\def\natexlab#1{#1}\fi

\bibitem[{{Bartczak} \& {Breiter}(2003)}]{BB03}
{Bartczak}, P., \& {Breiter}, S. 2003, Celestial Mechanics and Dynamical
  Astronomy, 86, 131

\bibitem[{{Batygin} \& {Morbidelli}(2015)}]{BM15}
{Batygin}, K., \& {Morbidelli}, A. 2015, \apj, 810, 110

\bibitem[{Belton {et~al.}(1996)Belton, Chapman, Klaasen, Harch, Thomas,
  Veverka, McEwen, \& Pappalardo}]{belton96}
Belton, M.~J., Chapman, C.~R., Klaasen, K.~P., {et~al.} 1996, \icarus, 120, 1

\bibitem[{{Belton} {et~al.}(1995){Belton}, {Chapman}, {Thomas}, {Davies},
  {Greenberg}, {Klaasen}, {Byrnes}, {D'Amario}, {Synnott}, {Johnson}, {McEwen},
  {Merline}, {Davis}, {Petit}, {Storrs}, {Veverka}, \& {Zellner}}]{belton95}
{Belton}, M.~J.~S., {Chapman}, C.~R., {Thomas}, P.~C., {et~al.} 1995, \nat,
  374, 785

\bibitem[{Benvenuto {et~al.}(1994)Benvenuto, Casati, \&
  Shepelyansky}]{benvenuto94}
Benvenuto, F., Casati, G., \& Shepelyansky, D.~L. 1994, Phys. Rev. Lett., 72,
  1818

\bibitem[{{Casati} {et~al.}(1988){Casati}, {Guarneri}, \&
  {Shepeliansky}}]{casati88}
{Casati}, G., {Guarneri}, I., \& {Shepeliansky}, D.~L. 1988, IEEE Journal of
  Quantum Electronics, 24, 1420

\bibitem[{{Chauvineau} {et~al.}(1993){Chauvineau}, {Farinella}, \&
  {Mignard}}]{CFM93}
{Chauvineau}, B., {Farinella}, P., \& {Mignard}, F. 1993, \icarus, 105, 350

\bibitem[{Chirikov(1979)}]{chirikov79}
Chirikov, B.~V. 1979, Physics Reports, 52, 263

\bibitem[{{Chirikov} \& {Vecheslavov}(1989)}]{chirikov89}
{Chirikov}, B.~V., \& {Vecheslavov}, V.~V. 1989, \aap, 221, 146

\bibitem[{{Emelyanenko}(1990)}]{emelyanenko90}
{Emelyanenko}, V.~V. 1990, Soviet Astronomy Letters, 16, 318

\bibitem[{Feng {et~al.}(2017)Feng, Noomen, Hou, Visser, \& Yuan}]{feng17}
Feng, J., Noomen, R., Hou, X., Visser, P., \& Yuan, J. 2017, Celestial
  Mechanics and Dynamical Astronomy, 127, 67

\bibitem[{{Feng} {et~al.}(2016){Feng}, {Noomen}, {Visser}, \& {Yuan}}]{FNV16}
{Feng}, J., {Noomen}, R., {Visser}, P., \& {Yuan}, J. 2016, Advances in Space
  Research, 58, 387

\bibitem[{Fujiwara {et~al.}(2006)Fujiwara, Kawaguchi, Yeomans, Abe, Mukai,
  Okada, Saito, Yano, Yoshikawa, Scheeres, Barnouin-Jha, Cheng, Demura,
  Gaskell, Hirata, Ikeda, Kominato, Miyamoto, Nakamura, Nakamura, Sasaki, \&
  Uesugi}]{fujiwara06}
Fujiwara, A., Kawaguchi, J., Yeomans, D.~K., {et~al.} 2006, Science, 312, 1330

\bibitem[{Fuse {et~al.}(2008)Fuse, Yoshida, Tholen, Ishiguro, \&
  Saito}]{fuse08}
Fuse, T., Yoshida, F., Tholen, D., Ishiguro, M., \& Saito, J. 2008, Earth,
  Planets and Space, 60, 33

\bibitem[{{Gaskell} {et~al.}(2008){Gaskell}, {Saito}, {Ishiguro}, {Kubota},
  {Hashimoto}, {Hirata}, {Abe}, {Barnouin-Jha}, \& {Scheeres}}]{gaskell08}
{Gaskell}, R., {Saito}, J., {Ishiguro}, M., {et~al.} 2008, NASA Planetary Data
  System, 92

\bibitem[{Heggie(1975)}]{heggie75}
Heggie, D.~C. 1975, \mnras, 173, 729

\bibitem[{Hergenrother \& Whiteley(2011)}]{hergenrother11}
Hergenrother, C.~W., \& Whiteley, R.~J. 2011, \icarus, 214, 194

\bibitem[{{Hu} \& {Scheeres}(2004)}]{HS04}
{Hu}, W., \& {Scheeres}, D.~J. 2004, \planss, 52, 685

\bibitem[{{Hut}(1981)}]{H81}
{Hut}, P. 1981, \aap, 99, 126

\bibitem[{Jorda {et~al.}(2016)Jorda, Gaskell, Capanna, Hviid, Lamy, Ďurech,
  Faury, Groussin, Gutiérrez, Jackman, Keihm, Keller, Knollenberg, Kührt,
  Marchi, Mottola, Palmer, Schloerb, Sierks, Vincent, A’Hearn, Barbieri,
  Rodrigo, Koschny, Rickman, Barucci, Bertaux, Bertini, Cremonese, Deppo,
  Davidsson, Debei, Cecco, Fornasier, Fulle, Güttler, Ip, Kramm, Küppers,
  Lara, Lazzarin, Moreno, Marzari, Naletto, Oklay, Thomas, Tubiana, \&
  Wenzel}]{jorda16}
Jorda, L., Gaskell, R., Capanna, C., {et~al.} 2016, \icarus, 277, 257

\bibitem[{{Kaasalainen, M.} {et~al.}(2003){Kaasalainen, M.}, {ki, T. Kwiatkow},
  {Abe, M.}, {Piironen, J.}, {Nakamura, T.}, {Ohba, Y.}, {Dermawan, B.},
  {Farnham, T.}, {Colas, F.}, {Lowry, S.}, {man, P. Wei}, {Whiteley, R. J.},
  {Tholen, D. J.}, {on, S. M. Lar}, {hikawa, M. Yo}, {Toth, I.}, \& {Velichko,
  F. P.}}]{alainen03}
{Kaasalainen, M.}, {ki, T. Kwiatkow}, {Abe, M.}, {et~al.} 2003, \aap, 405, L29

\bibitem[{Lages \& Shepelyansky(2013)}]{lages13}
Lages, J., \& Shepelyansky, D.~L. 2013, \mnras: Letters, 430, L25

\bibitem[{{Lichtenberg} \& {Lieberman}(1992)}]{lichtenberg92}
{Lichtenberg}, A.~J., \& {Lieberman}, M.~A. 1992, Regular and Chaotic Dynamics
  (Springer New York)

\bibitem[{{Lowry, S. C.} {et~al.}(2014){Lowry, S. C.}, {Weissman, P. R.},
  {Duddy, S. R.}, {Rozitis, B.}, {Fitzsimmons, A.}, {Green, S. F.}, {Hicks, M.
  D.}, {Snodgrass, C.}, {Wolters, S. D.}, {Chesley, S. R.}, {Pittichová, J.},
  \& {van Oers, P.}}]{lowry14}
{Lowry, S. C.}, {Weissman, P. R.}, {Duddy, S. R.}, {et~al.} 2014, \aap, 562,
  A48

\bibitem[{Malyshkin \& Tremaine(1999)}]{malyshkin99}
Malyshkin, L., \& Tremaine, S. 1999, \icarus, 141, 341

\bibitem[{{Marchis} {et~al.}(2014){Marchis}, {Durech}, {Castillo-Rogez},
  {Vachier}, {Cuk}, {Berthier}, {Wong}, {Kalas}, {Duchene}, {van Dam},
  {Hamanowa}, \& {Viikinkoski}}]{MDC14}
{Marchis}, F., {Durech}, J., {Castillo-Rogez}, J., {et~al.} 2014, \apjl, 783,
  L37

\bibitem[{Meiss(1992)}]{meiss92}
Meiss, J.~D. 1992, Rev. Mod. Phys., 64, 795

\bibitem[{Melnikov \& Shevchenko(2010)}]{melnikov10}
Melnikov, A., \& Shevchenko, I. 2010, \icarus, 209, 786

\bibitem[{{Mysen} \& {Aksnes}(2007)}]{MA07}
{Mysen}, E., \& {Aksnes}, K. 2007, \aap, 470, 1193

\bibitem[{{Mysen} {et~al.}(2006){Mysen}, {Olsen}, \& {Aksnes}}]{MOA06}
{Mysen}, E., {Olsen}, {\O}., \& {Aksnes}, K. 2006, \planss, 54, 750

\bibitem[{{Olsen}(2006)}]{O06}
{Olsen}, {\O}. 2006, \aap, 449, 821

\bibitem[{Petit {et~al.}(1997)Petit, Durda, Greenberg, Hurford, \&
  Geissler}]{petit97}
Petit, J., Durda, D., Greenberg, R., Hurford, T., \& Geissler, P. 1997,
  \icarus, 130, 177

\bibitem[{Petrosky(1986)}]{petrosky86}
Petrosky, T.~Y. 1986, Physics Letters A, 117, 328

\bibitem[{Popova \& Shevchenko(2016)}]{popova16}
Popova, E.~A., \& Shevchenko, I.~I. 2016, Astronomy Letters, 42, 474

\bibitem[{Pravec {et~al.}(2008)Pravec, Harris, Vokrouhlick{\'y}, Warner,
  Ku{\v{s}}nir{\'a}k, Hornoch, Pray, Higgins, Oey, Gal{\'a}d, \v{S}.
  Gajdo{\v{s}}, Korno{\v{s}}, Vil{\'a}gi, Hus{\'a}rik, Krugly, Shevchenko,
  Chiorny, Gaftonyuk, Jr., Gross, Terrell, Stephens, Dyvig, Reddy, Ries, Colas,
  Lecacheux, Durkee, Masi, Koff, \& Goncalves}]{pravec08}
Pravec, P., Harris, A., Vokrouhlick{\'y}, D., {et~al.} 2008, \icarus, 197, 497

\bibitem[{{Quillen} {et~al.}(2017){Quillen}, {Nichols-Fleming}, {Chen}, \&
  {Noyelles}}]{Q17}
{Quillen}, A.~C., {Nichols-Fleming}, F., {Chen}, Y.-Y., \& {Noyelles}, B. 2017,
  ArXiv e-prints, arXiv:1701.05594

\bibitem[{Rollin {et~al.}(2015)Rollin, Haag, \& Lages}]{rollin15b}
Rollin, G., Haag, P., \& Lages, J. 2015, Physics Letters A, 379, 1017

\bibitem[{{Rollin} {et~al.}(2015){Rollin}, {Lages}, \&
  {Shepelyansky}}]{rollin15a}
{Rollin}, G., {Lages}, J., \& {Shepelyansky}, D.~L. 2015, \aap, 576, A40

\bibitem[{Roy \& Haddow(2003)}]{roy03}
Roy, A., \& Haddow, M. 2003, Celestial Mechanics and Dynamical Astronomy, 87,
  411

\bibitem[{Scheeres(2007)}]{scheeres07}
Scheeres, D. 2007, \icarus, 189, 370

\bibitem[{{Scheeres}(1994)}]{Sch94}
{Scheeres}, D.~J. 1994, \icarus, 110, 225

\bibitem[{{Scheeres}(2012)}]{Sch12}
---. 2012, Acta Astronautica, 72, 1

\bibitem[{{Scheeres} {et~al.}(1996){Scheeres}, {Ostro}, {Hudson}, \&
  {Werner}}]{SOH96}
{Scheeres}, D.~J., {Ostro}, S.~J., {Hudson}, R.~S., \& {Werner}, R.~A. 1996,
  \icarus, 121, 67

\bibitem[{{Scheeres} {et~al.}(2000){Scheeres}, {Williams}, \& {Miller}}]{SWM00}
{Scheeres}, D.~J., {Williams}, B.~G., \& {Miller}, J.~K. 2000, Journal of
  Guidance Control Dynamics, 23, 466

\bibitem[{Shevchenko(2010)}]{shevchenko10}
Shevchenko, I.~I. 2010, Phys. Rev. E, 81, 066216

\bibitem[{Shevchenko(2011)}]{shevchenko11}
---. 2011, \na, 16, 94

\bibitem[{Shevchenko(2015)}]{shevchenko15}
---. 2015, \apj, 799, 8

\bibitem[{{Vokrouhlick{\'y}} {et~al.}(2003){Vokrouhlick{\'y}}, {Nesvorn{\'y}},
  \& {Bottke}}]{vokrouhlick03}
{Vokrouhlick{\'y}}, D., {Nesvorn{\'y}}, D., \& {Bottke}, W.~F. 2003, \nat, 425,
  147

\bibitem[{Warner {et~al.}(2009)Warner, Harris, \& Pravec}]{warner09}
Warner, B.~D., Harris, A.~W., \& Pravec, P. 2009, \icarus, 202, 134

\bibitem[{{Werner}(1994)}]{W94}
{Werner}, R.~A. 1994, Celestial Mechanics and Dynamical Astronomy, 59, 253

\bibitem[{{Werner} \& {Scheeres}(1996)}]{WS96}
{Werner}, R.~A., \& {Scheeres}, D.~J. 1996, Celestial Mechanics and Dynamical
  Astronomy, 65, 313

\bibitem[{{Whiteley} {et~al.}(2002){Whiteley}, {Hergenrother}, \&
  {Tholen}}]{whiteley02}
{Whiteley}, R.~J., {Hergenrother}, C.~W., \& {Tholen}, D.~J. 2002, in ESA
  Special Publication, Vol. 500, Asteroids, Comets, and Meteors: ACM 2002, ed.
  B.~{Warmbein}, 473--480

\bibitem[{{Yu} \& {Baoyin}(2012)}]{YB12}
{Yu}, Y., \& {Baoyin}, H. 2012, \aj, 143, 62

\bibitem[{{Zimmer} {et~al.}(2014){Zimmer}, {Williams}, {Johnson}, {Church},
  {Fevig}, \& {Semke}}]{zimmer14}
{Zimmer}, M., {Williams}, K., {Johnson}, J., {et~al.} 2014, in Lunar and
  Planetary Science Conference, Vol.~45, 2226

\end{thebibliography}
\bibliographystyle{aasjournal}

\end{document}